\documentclass[reprint, superscriptaddress, nofootinbib, amsmath, amssymb, aps]{revtex4-1}

\usepackage{graphicx}
\usepackage{dcolumn}
\usepackage{verbatim}
\usepackage{bm}
\usepackage{todonotes}
\usepackage{lmodern}
\usepackage{tikz}
\usepackage{pgfplots}
\pgfplotsset{compat=newest,every axis plot/.append style={line width=1pt}}
\usepackage[colorlinks,linktocpage,linkcolor=cyan,citecolor=cyan]{hyperref}
\usepackage{tabularx}
\definecolor{lightgray}{gray}{0.9}
\definecolor{Amber}{rgb}{1.0, 0.75, 0.0}
\definecolor{blizzardblue}{rgb}{0.67, 0.9, 0.93}

\begin{document}

\title{Lattice Simulation of Multi-Stream Inflation}

\author{Tingqi Cai}
\email{tq30701@mail.ustc.edu.cn}
\affiliation{Department of Astronomy, School of Physical Sciences, University of Science and Technology of China, Hefei, Anhui 230026, China}
\affiliation{School of Astronomy and Space Science, University of Science and Technology of China, Hefei, Anhui 230026, China}
\affiliation{CAS Key Laboratory for Researches in Galaxies and Cosmology, University of Science and Technology of China, Hefei, Anhui 230026, China}

\author{Jie Jiang}
\email{jiejiang@mail.ustc.edu.cn}
\affiliation{Department of Astronomy, School of Physical Sciences, University of Science and Technology of China, Hefei, Anhui 230026, China}
\affiliation{School of Astronomy and Space Science, University of Science and Technology of China, Hefei, Anhui 230026, China}
\affiliation{CAS Key Laboratory for Researches in Galaxies and Cosmology, University of Science and Technology of China, Hefei, Anhui 230026, China}

\author{Yi Wang}
\email{phyw@ust.hk}
\affiliation{Department of Physics, The Hong Kong University of Science and Technology,\\
Clear Water Bay, Kowloon, Hong Kong, P.R.China}
\affiliation{Jokey Club Institute for Advanced Study, The Hong Kong University of Science and Technology,\\
Clear Water Bay, Kowloon, Hong Kong, P.R.China}


\begin{abstract}
We present the first lattice simulation to investigate the nature of multi-stream inflation. The simulation confirms the physical picture of multi-stream inflation, and with new findings in parameter space and field behaviors. Our simulation shows that gradient energy plays a significant role in multi-stream inflation. For a double field potential with a shifted Gaussian barrier, bifurcation probability is controlled by the shift distance with an error function relation. The bubbles created by bifurcation tend to be more spherical as bifurcation probability decreases. Also, the bifurcation is more likely to introduce oscillations of field values inside the bubbles than outside.
\end{abstract}


\maketitle


\section{introduction}

Inflation has been the leading paradigm of the early universe \cite{Guth:1980zm, Linde:1981mu, Starobinsky:1980te, Brout:1977ix, Sato:1980yn, Fang:1980wi, Albrecht:1982wi} to solve several conundrums of the Hot Big Bang, and provide origin of the density fluctuations which serve as the seeds of large scale structure. 

\begin{figure}[ht]
    \centering
	\includegraphics[width=0.46\textwidth]{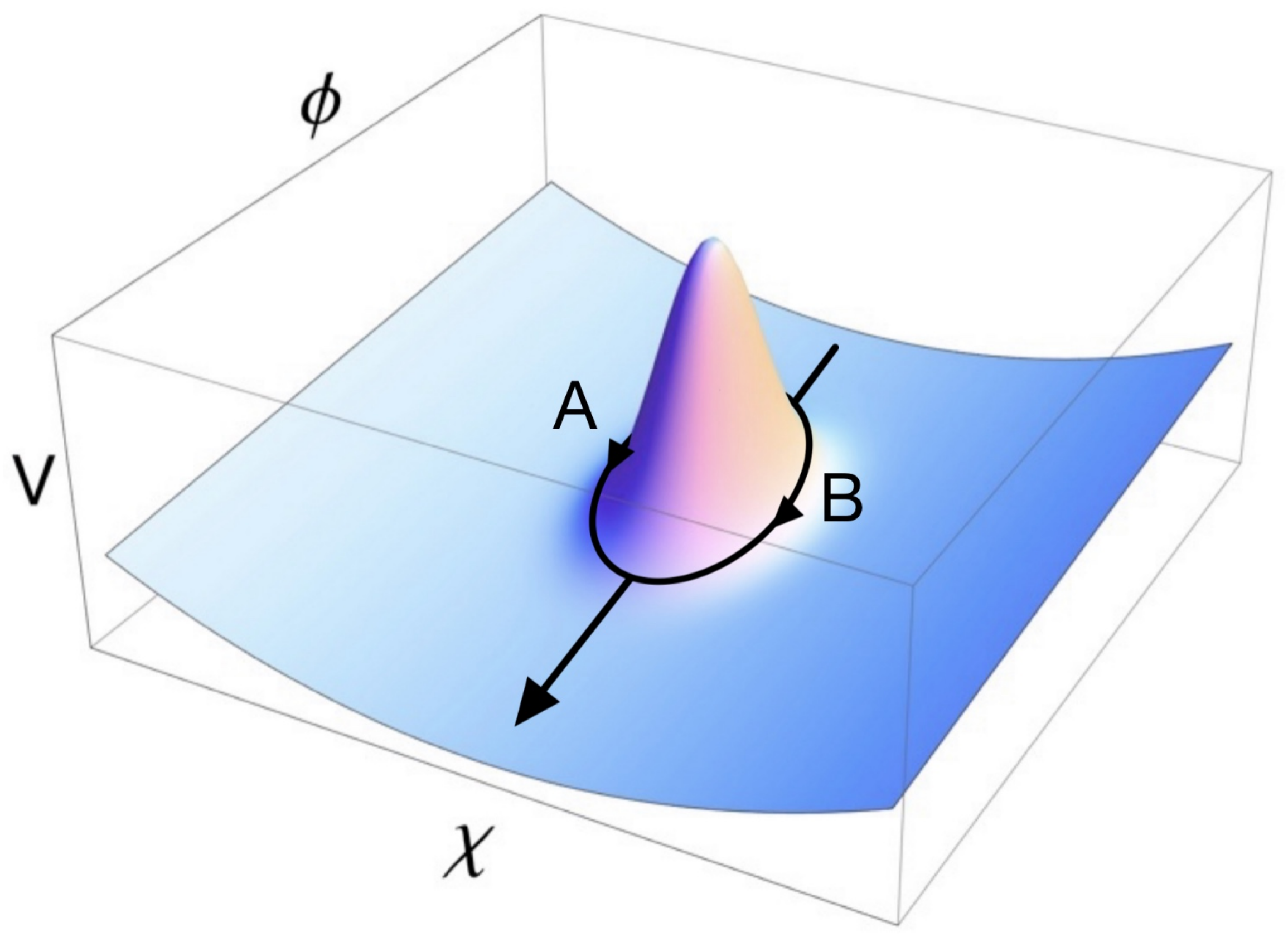}
	\caption{Multi-stream inflation: how the bifurcation in the isocurvature direction during inflation generate bubbles, inflaton for most part of the universe follows trajectory A while some follow trajectory B.}
	\label{f0}
\end{figure}

Motivated by the string landscape \cite{Bousso:2000xa, Susskind:2003kw}, the inflationary dynamics may be not as simple as the prototype. As illustrated in Fig.~\ref{f0}, in the multi-field space, if the inflationary trajectory encounters a barrier, bifurcation of the inflationary trajectory can happen. The local universes may undergo inflation along either A or B trajectories as marked in the figure. This bifurcation behavior is known as multi-stream inflation \cite{Li:2009sp}. Bifurcation behavior in a random potential is spotted in \cite{Duplessis:2012nb} and increasing the dimension of the random field space can increase the probability of bifurcations \cite{Liu:2015dda}.

The observational consequences of multi-stream inflation include: (i) Due to the $\delta N$ formalism \cite{Starobinsky:1985ibc,Sasaki:1995aw,Lyth:2005fi}, the e-folding number difference between multiple trajectories is an additional source of curvature perturbation at the bifurcation scale \cite{Li:2009sp}. Depending on the scale of bifurcation, this additional perturbation may be responsible for features on the CMB, the CMB cold spot \cite{Afshordi:2010wn}, or primordial black hole formation \cite{Nakama:2016kfq} when the comoving scale corresponding to the bifurcation returns to the horizon. (ii) The local environments along different inflationary trajectories (A and B in Fig.~\ref{f0}) can be different. The difference can be responsible for position space anomalies of the universe, such as power asymmetries on the CMB \cite{Li:2009sp,Wang:2013vxa}, clustered production of primordial black holes \cite{Ding:2019tjk} and stellar bubbles \cite{Cai:2021zxo}. And correlation between the bifurcation scale and the local environments introduce non-Gaussianities \cite{Li:2009sp, Abolhasani:2010kn}. (iii) The domain wall in between A and B can have observational effects \cite{Li:2009me}, and the thickness of the domain wall can help evade the Sunyaev-Zeldovich effect constraints when using voids to ease the Hubble tension problem \cite{Ding:2019mmw}. If the bifurcated trajectories finally combine (or merge to connected degenerate vacuum configurations in the case of multiple isocurvature directions), the tension of the domain wall finally vanish and thus it does not cause the conventional domain wall problem in cosmology.

Although many observational features in multi-stream inflation can be estimated by the separate universe picture and the $\delta N$ formalism, quantitative analytical calculation is usually difficult in multi-stream inflation. This is because of the temporary domain wall in between the two trajectories. The domain wall is dynamical and introduces additional gradient energy to the field configuration. Although the observational consequences of the domain wall is discussed in the literature \cite{Li:2009me, Ding:2019mmw}, how the domain wall affects the bifurcation dynamics was not studied. Also, the previous numerical simulations of the bifurcation probability were built upon the $\delta N$ formalism which did not take the domain wall and gradient energy into account \cite{Duplessis:2012nb, Liu:2015dda}. 
Statobinsky's stochastic approach was used to study the dynamics of bifurcation \cite{Liu:2015dda}. The difference between its analysis of bifurcation and that of lattice codes is that stochastic approach can only show the fluctuations away from the domain wall, but not the ones close to it due to the strong gradient and potential energy. 
Since the interplay between the domain wall and the nearby bifurcated field is nonlinear, and inhomogeneous in space, we expect that the effect has to be simulated by lattice simulation. Also, the bifurcation may introduce oscillations, which is also strongly affected by the gradient energy.

In the literature of the very early universe cosmology, lattice codes are widely used in simulating reheating after inflation. During inflation, simulations are possible (see, for example, \cite{Caravano:2021pgc}) but the computation is challenging because of the exponential expansion of the space volume. In our case, we will by-pass this difficulty by studying only a few e-folds of inflation around the bifurcation time. This suffices the study of the dynamics of the domain wall and how it affects the bifurcation. In this paper, we use a modified version of LATTICEEASY \cite{Felder:2000hq} to simulate the multi-stream inflation.

This paper is organized as follows. Sec.~\ref{0} provides the basic equations for the dynamics of inflation. Then we take a potential of multi-stream inflation, and simulate bifurcations in Sec.~\ref{3}. We discuss the effects of gradient energy, then investigate some important features of multi-stream inflation, including bifurcation probability, potential shift, bubble sphericity as well as oscillatory behaviors of trajectories corresponding to the bubble region. We conclude in Sec.~\ref{7}. Some details about the calculation of background potential and bubble sphericity are presented in Appendix~\ref{2} and \ref{a1}.

We will work in Planck units, with $c=\hbar=G=1$ and the reduced Planck mass $M_p=1/\sqrt{8\pi}$. Dot is used to denote derivative with respect to cosmic time ($\dot{\ }=d/dt $).

\section{\label{0}Evolution Equations}

We study the dynamics of scalar fields, ignoring the backreaction by gravity. We start from the Friedmann-Lema\^{i}tre-Robertson-Walker (FLRW) metric:
\begin{equation}
	ds^2=-dt^2 + a^2(t)(dx^2+dy^2+dz^2), 
\end{equation}
where $ a $ is the scale factor. And the Friedmann equations for background evolution are:
\begin{equation}
	3H^2=8\pi\rho,\ \ \frac{\ddot a}{a}=-\frac{4\pi}{3}(\rho+3p),
\end{equation}
where $ H \equiv \dot{a } / a $ is the Hubble parameter describing the expansion rate of the universe. The energy density $ \rho $ and pressure $ p $ for scalar fields are
\begin{equation}
	\rho=T+G+V,\ \ p=T-\frac{1}{3}G-V,
\end{equation}
with kinetic energy $ T $ and gradient energy $ G $:
\begin{equation}
	T = \frac{1}{2} \dot{\phi} ^ 2 + \frac{1}{2} \dot{\chi} ^ 2,\ \ G = \frac{1}{2a^2} |\nabla\phi| ^ 2 + \frac{1}{2a^2} |\nabla\chi| ^ 2.
\end{equation}
The equations of motion for the scalar fields are
\begin{equation}
	\begin{split}
		\ddot\phi+3H\dot\phi-\frac{\nabla^2}{a^2}\phi+\frac{\partial V}{\partial\phi}=0,\\
		\ddot\chi+3H\dot\chi-\frac{\nabla^2}{a^2}\chi+\frac{\partial V}{\partial\chi}=0.
	\end{split}
\end{equation}

Different from LATTICEEASY, we here uses Runge-Kutta methods to numerically solve these classical equations. We focus on the evolution of such inhomogeneous background fields in our parameter space. As for the quantum nature of the inflaton, it is described by the statistical uncertainty generated by semi-classical approximation \cite{Felder:2000hq,Caravano:2021pgc}. That is to say the methods we use to determine initial conditions are the same as LATTICEEASY.

\begin{figure}
    \centering
	\includegraphics[width=0.46\textwidth]{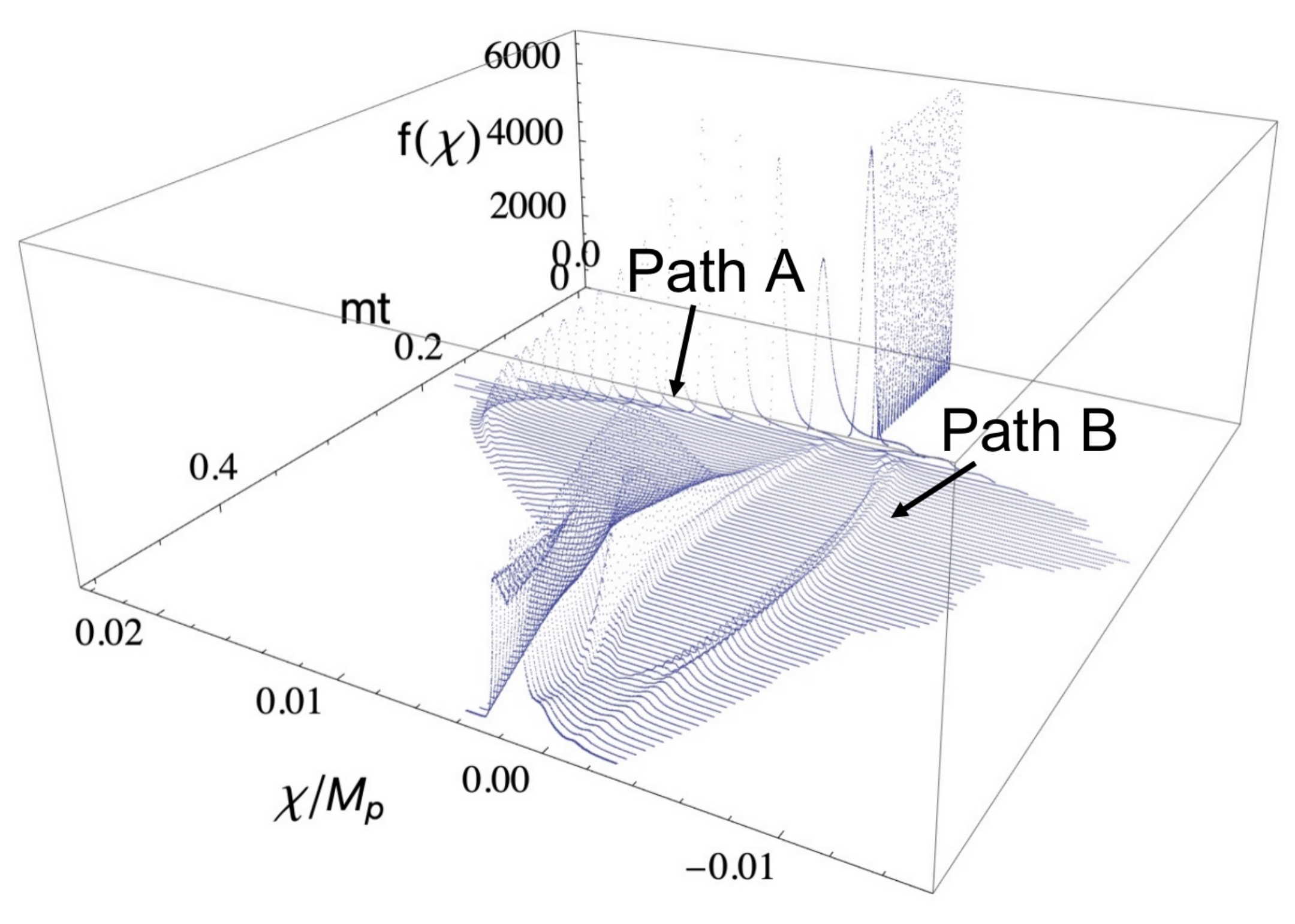}
	\caption{Simulated inflaton trajectories in the potential with a barrier. $f(\chi)$ represents probability density of $\chi$ at different time. Trajectory A rolls on one side of the barrier and B rolls on the other side of it after bifurcation and before combination of these trajectories.}
	\label{f2}
\end{figure}

\begin{figure}
    \centering
	\includegraphics[width=0.46\textwidth]{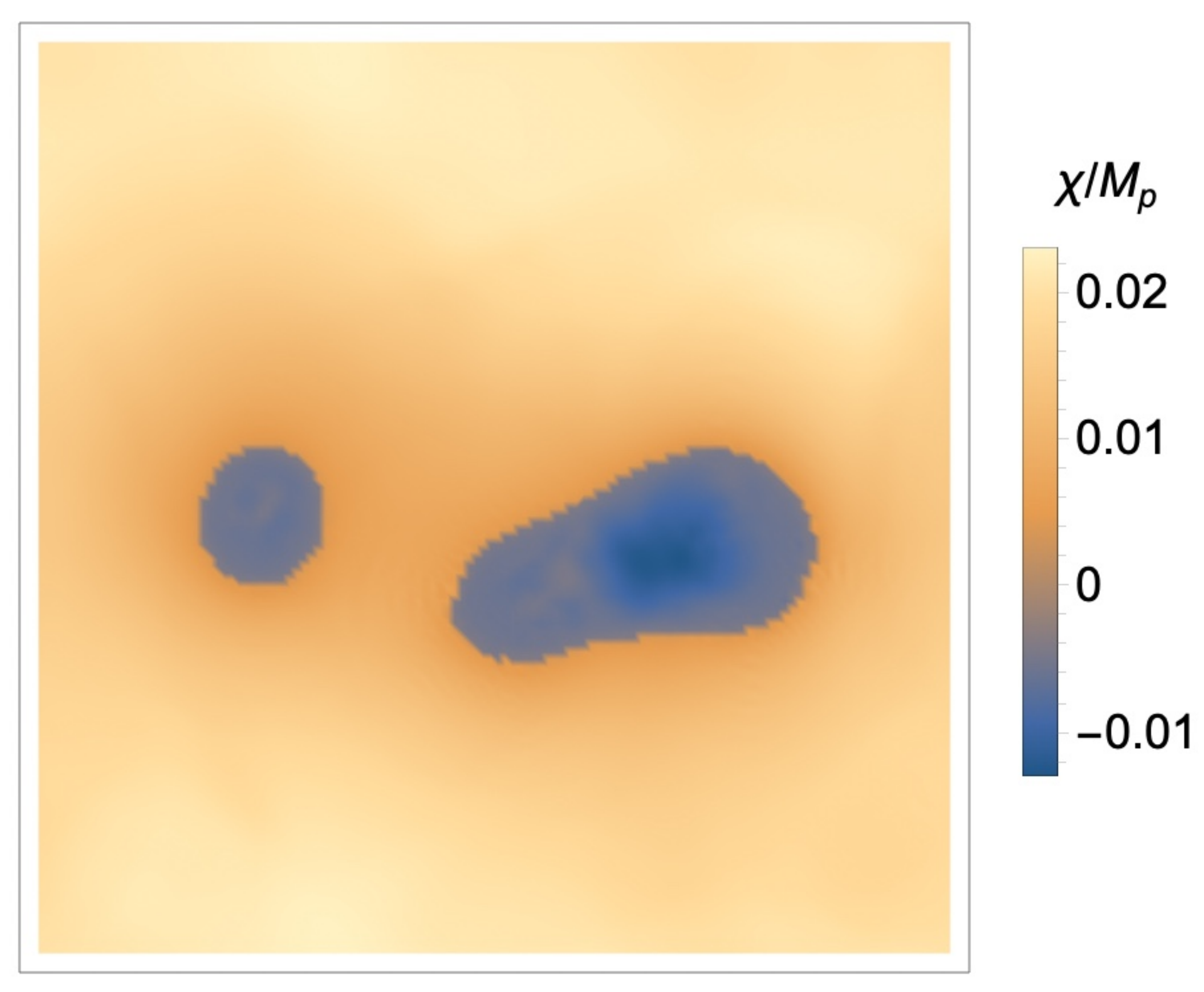}
	\caption{Two-dimensional slice of the lattice points during the bifurcation, we can see the cross-sections of the bubbles created, which are the blue parts.}
	\label{f44}
\end{figure}

\section{\label{3}Toy Model with a barrier}

We consider a potential with a barrier, around which multi-stream inflation takes place:
\begin{equation}
    \label{e1}
	\begin{split}
		V(\phi,\chi&)=\frac{1}{2}m^2\phi^2+\frac{1}{2}\alpha m^2\chi^2+ \\
		& \beta m^2\text e^{-[\frac{(\phi-\phi_b)^2}{2\sigma_1^2}+\frac{(\chi-\chi_b)^2}{2\sigma_2^2}]},
	\end{split}
\end{equation}
where $\beta$ is a constant controlling the height of the barrier. The parameters  $\phi_b, \chi_b$ decides barrier position and $\phi_0=16M_p,\chi_0=0$ are the initial values of fields.

As shown in Fig.~\ref{f0} and Fig.~\ref{f44} the inflaton rolls down this potential, its trajectories bifurcate after hitting the barrier, creating bubbles in the universe. We then focus on some phenomena concerning such bifurcation. Fig.~\ref{f2} is a series of plots of trajectory $\chi$ distributions, where we can easily see the whole process of bifurcation.

\subsection{\label{35}Gradient Energy}

In this subsection we show the effect of gradient energy, and how it affects the dynamics of bifurcation. If we set the potential of the barrier to be a bit shifted from the center of the background potential, that is to set $0<\chi_b\ll1$ in Eq.~\ref{e1}, then we have an asymmetric bifurcation as an example condition for simulation.

\begin{figure}[ht]
    \centering
	\includegraphics[width=0.48\textwidth]{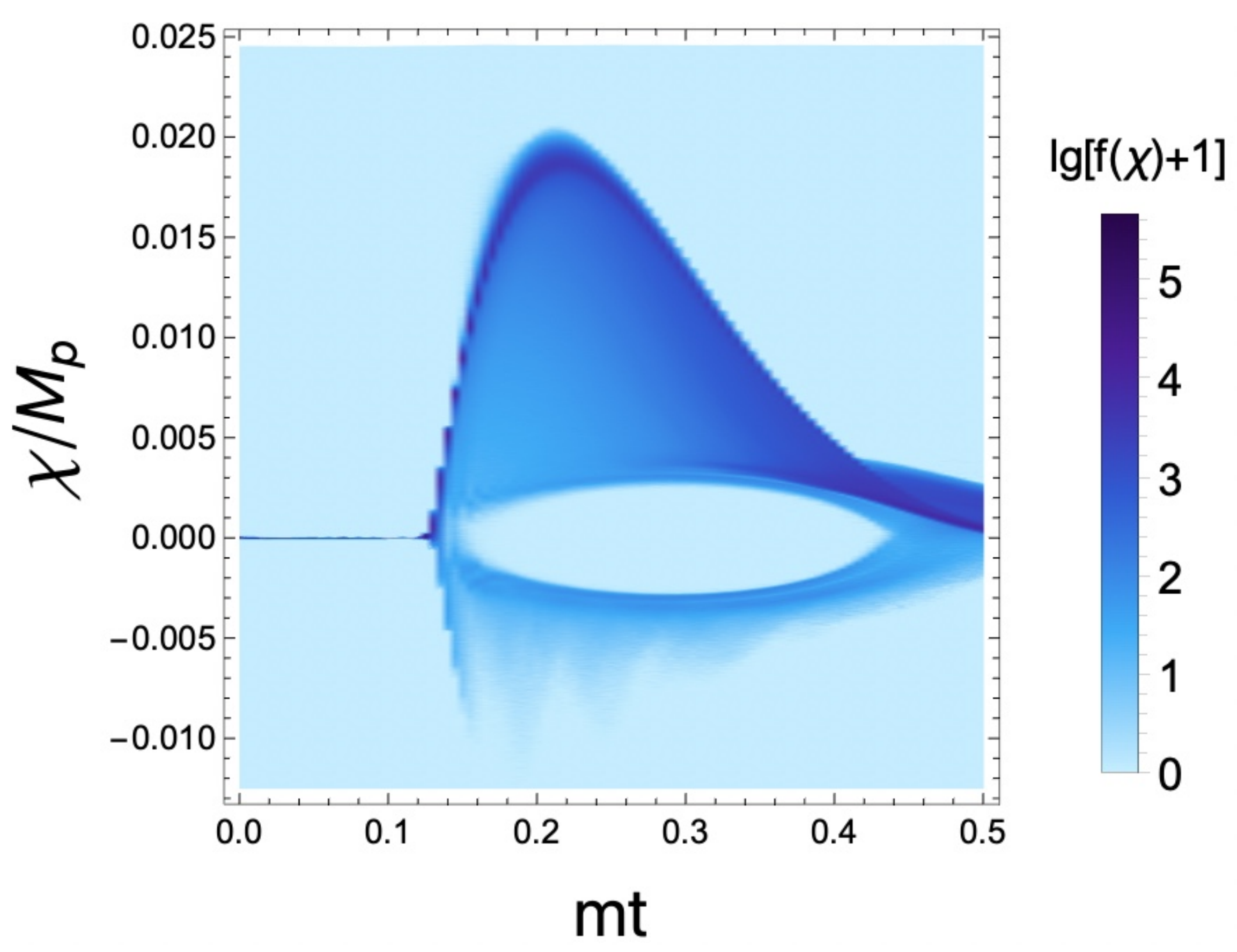}
	\caption{$\chi$ value distribution of trajectories in different time, gradient energy is included in the simulation as it should be. $f(\chi)$ is the probability distribution.}
	\label{f7}
\end{figure}

In this simulation above, the distribution of trajectories is shown as Fig.~\ref{f7}. While a small number of trajectories roll below the potential barrier, most of them roll above the barrier, these two kinds of trajectories correspond to trajectory B and A in Fig.~\ref{f2} respectively.

\begin{figure}[ht]
    \centering
	\includegraphics[width=0.48\textwidth]{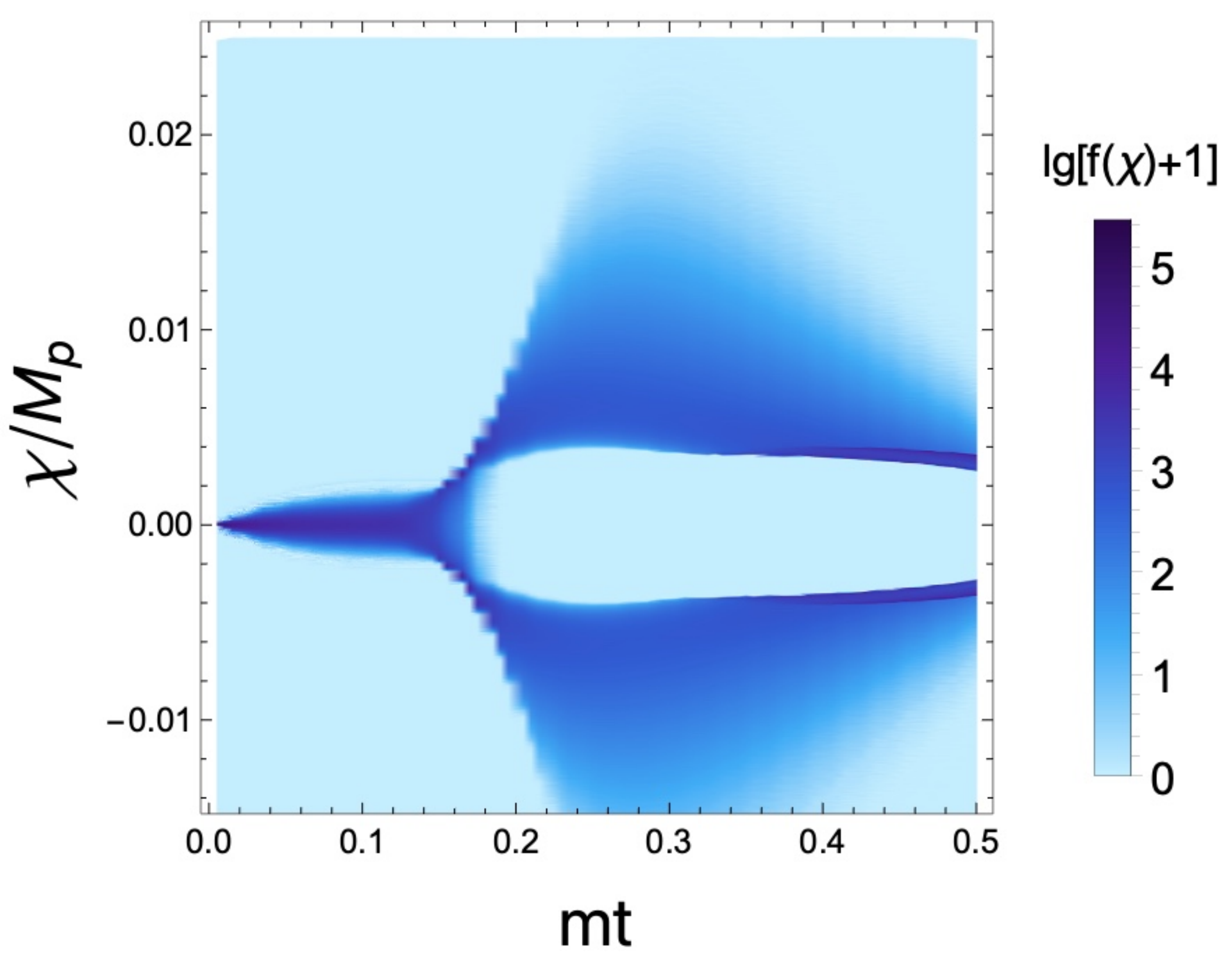}
	\caption{$\chi$ value distribution of trajectories in different time with gradient energy being set to zero. $f(\chi)$ is the probability density of distribution.}
	\label{f35}
\end{figure}

We can immediately realize that although there are both bifurcations happening in Fig.~\ref{f2} and Fig.~\ref{f7}, there are obvious differences in the shape of trajectories. In Fig.~\ref{f2} we assume the trajectories would stay close to the barrier just as the black curves show. However, in Fig.~\ref{f7} where the actual trajectories are shown, it turns out that most of the trajectories roll away from the barrier before coming back due to the $\alpha m^2\chi^2/2$ term. Does gradient energy contribute to such behavior? In order to find out we run the simulation again using all the same conditions but setting the gradient energy to zero. The result of such simulation is shown as Fig.~\ref{f35}. Without gradient energy, the lattice points in the simulation box are independent of each other and the shapes of trajectories are fully determined by the potential, leading to the fact that before bifurcation the distribution in Fig.~\ref{f35} is wider and more symmetric than that in Fig.~\ref{f7}. As the distributions have many differences with or without gradient energy, it surely has critical effect on the bifurcation itself. That is to say, gradient energy does contribute to the behavior of trajectories in a complicated way. It is hard to evaluate such effect analytically, so numerical simulation might be a more appropriate way.

The effects of gradient energy here is complicated, we cannot simply claim that its existence always help making the trajectories move away from the barrier, instead this is only confirmed in the parameter space that we choose. For example, what if the barrier in the potential is much larger, will the trajectories move closer from the barrier? We are unable to answer this question at the moment because the number of e-folds that lattice simulation can run is limited.

As we discuss above, the $\chi$ distribution of trajectories is wider when the potential energy is gone before bifurcation, that is because although initial value fluctuation of $\chi$ is relatively small at $(mt,\chi/M_p)=(0,0)$, trajectories will immediately move towards different directions and apart from each other since they have different initial field derivatives $\dot\chi$. 

Another issue is the bifurcation probability. In Fig.~\ref{f7} and Fig.~\ref{f35}, apart from the gradient energy, we have exactly identical initial conditions and potential barrier position $\phi_b,\chi_b$, yet the simulation shows that in Fig.~\ref{f35} the bifurcation is nearly symmetric, while in Fig.~\ref{f7}, asymmetry is significant and only a small fraction of trajectories take path B (downwards) instead of path A (upwards). Clearly it is the existence of gradient energy that cause the significant asymmetry. Gradient energy creates a strong bound among the trajectories. That is to say the trajectories tend to move as a group facing a bifurcation, this is confirmed by Fig.~\ref{f7} since we can see that trajectory distribution has been narrow. 

As gradient energy was not considered when $\delta N$ formalism was used to estimate some features in previous works, the results were not enough to recover many of the details during bifurcation. This is no longer a problem in this paper since lattice simulation can take gradient energy into account.

\subsection{\label{4}Bifurcation Probability and Potential Shift}

The bifurcation probability is influenced by the position of the barrier, if the barrier center is at the bottom of potential $\alpha m^2\chi^2/2$ term. In other words, if $\chi_b=\chi_0=0$ in Eq.~\ref{e1}, then the bifurcation is almost symmetric. When moving away from this symmetric point, the bifurcation probability becomes asymmetric. 

In the separate universe picture, we expect that the bifurcation probability is an error function as a function of $\chi_b$. This is because the bifurcation is controlled by the fluctuation of $\chi$ exceeding a threshold, which follows an approximately Gaussian distribution.

Considering the effect of gradient energy, the relation between the bifurcation probability and $\chi_b$ is less obvious. Because the gradient energy affects the bifurcation probability significantly. Nevertheless, our simulation result show that this relation still takes the form of an error function. 

\begin{figure}[ht]
    \centering
	\includegraphics[width=0.44\textwidth]{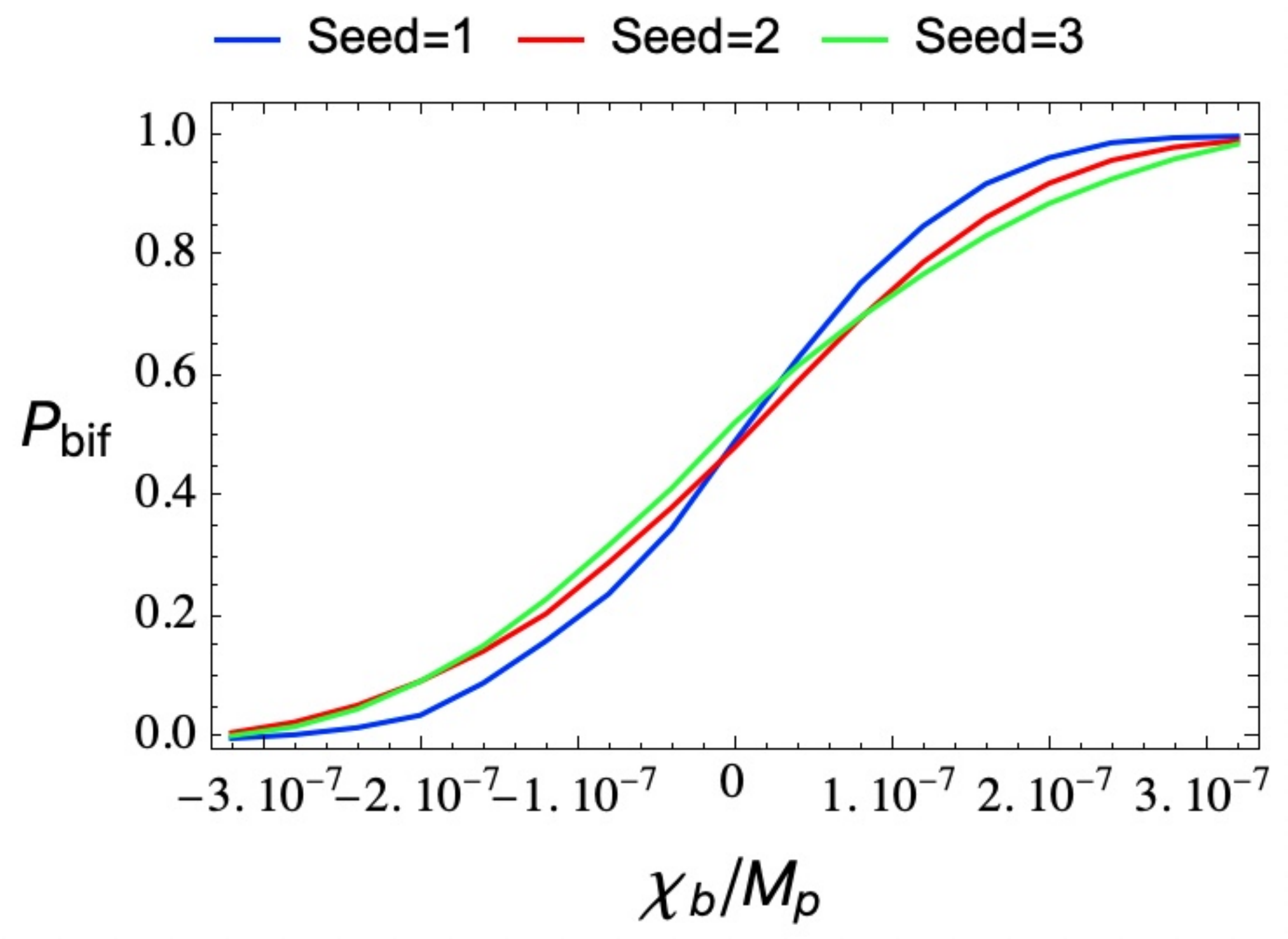}
	\caption{Simulation results of bifurcation probability as a function of shift distance, different colors represent different initial random seed. Randomness of initial fluctuation might cause small changes in the curves, but it does not change the fact that each curve is very close to error function curve.}
	\label{f3}
\end{figure}

\begin{figure}[ht]
    \centering
	\includegraphics[width=0.44\textwidth]{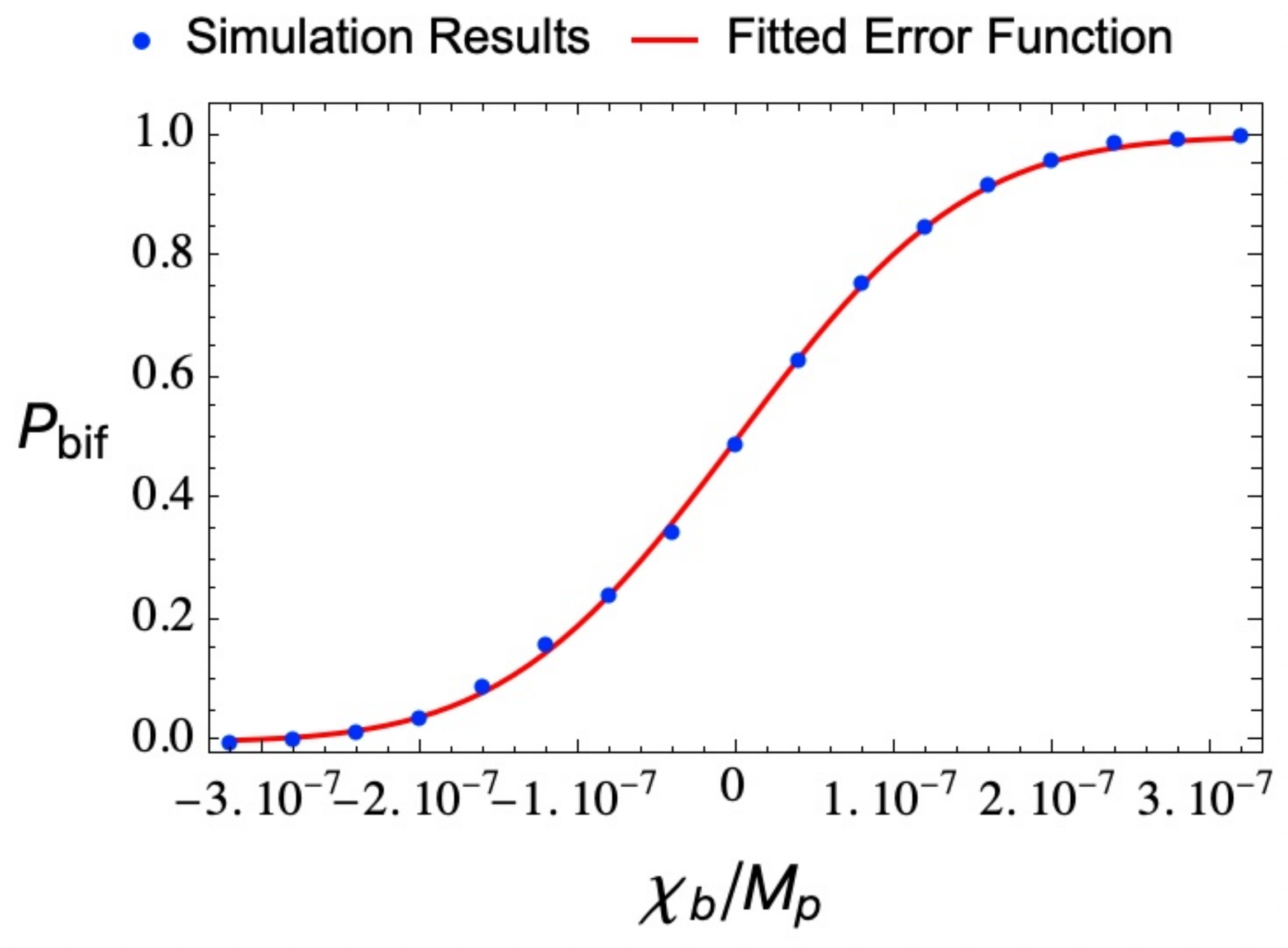}
	\caption{One example of simulation results and fitted curve, blue dots are the simulation outputs, yellow curve is the fitted function.}
	\label{f4}
\end{figure}

Fig.~\ref{f3} shows that for different random seeds, which means for different initial fluctuations, bifurcation probability increases smoothly as the potential barrier moves in $\chi$ direction. As displayed in Fig.~\ref{f4}, this function can be well fitted by the error function 
\begin{equation}
	\text{P}_{\text{bif}}=\frac{1}{2}+\frac{1}{2}\text{erf}(\frac{\delta}{\sqrt2\sigma}),
	\label{e8}
\end{equation}
where $\delta=\chi_b/M_p$ is the rescaled potential shift distance.

It is natural to try to estimate the bifurcation probability by dividing gaussian distributions of $\chi$ field using position of barrier center $\chi_b$. While this indeed can lead to an error function relation, the parameter $\sigma$ of such function is significantly larger than the one obtained from simulation. The reason of it has been discussed in Sec.~\ref{35}. As gradient energy tend to connect different trajectories, a slight shift of the barrier position can cause the trajectory distribution to move as a whole, consequently reducing the value of $\sigma$ in Eq.~\ref{e8}

\subsection{\label{5}Bubble Sphericity and Bifurcation Probability}

In lattice simulation we can extract bubbles and analyze their shapes. For very small bifurcation probability $\text{P}_{\text{bif}}\ll1$, the shapes of bubbles are expected to be spherical \cite{Afshordi:2010wn}, following the peak theory \cite{Bardeen:1985tr} (which will then be affected by the gradient energy of the field at the scale of the bifurcation), or alternatively considered as instantons in quantum field theory minimizing the Euclidean action \cite{Coleman:1977py}. It is thus theoretically expected for a bubble that the smaller $\text{P}_{\text{bif}}$, the more spherical.

However, if we are not talking about exponentially small $\text{P}_{\text{bif}}$, for small but still significant $\text{P}_{\text{bif}}$, the trend of sphericity is less clear. In this subsection, we test the sphericity as a function of $\text{P}_{\text{bif}}$. We confirm the trend of more spherical bubbles for a smaller bifurcation probability, but also show that the result has large variance, and the initial condition also affects the sphericity significantly. To eliminate the impact from the initial condition, we choose to focus on a certain simulation setup with fixed initial condition, while changing bifurcation probability by shifting the position of the barrier, to track how it changes sphericity. 

To measure how spherical a bubble is, limited by the resolution of the lattice, we compare the bubble with a spherical bubble with the same volume, and calculate their overlapping ratio. In this definition a perfectly spherical bubble has sphericity equals to 1. We tested that this definition is robust for a calculation on the lattice. The details and benchmarks of the definition of sphericity is provided in Appendix~\ref{a1}.

Fig.~\ref{f5} reveals that for different bubbles, although their sphericity curves have can be quite different, these curves all show that sphericity tends to be smaller as bifurcation probability increases. Here we have chosen the parameter space for intermediate bifurcation probability, to have enough samples of bubbles with clear bubble identification.

\begin{figure}
    \centering
	\includegraphics[width=0.44\textwidth]{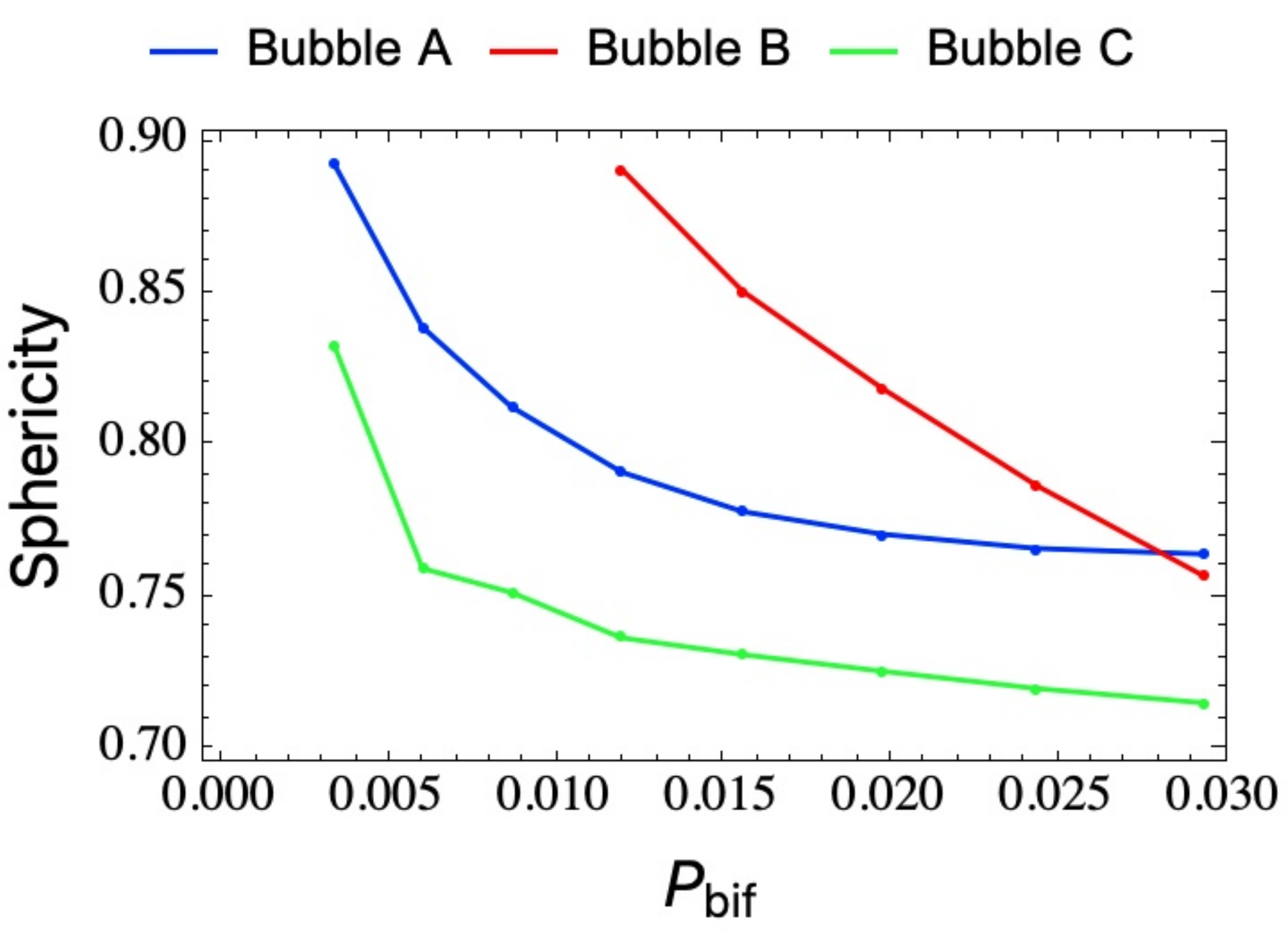}
	\caption{Sphericity as a function of bifurcation probability, different color represents different bubble.}
	\label{f5}
\end{figure}

\subsection{\label{6}Oscillation of the Individual Trajectories}

\begin{figure}
    \centering
	\includegraphics[width=0.44\textwidth]{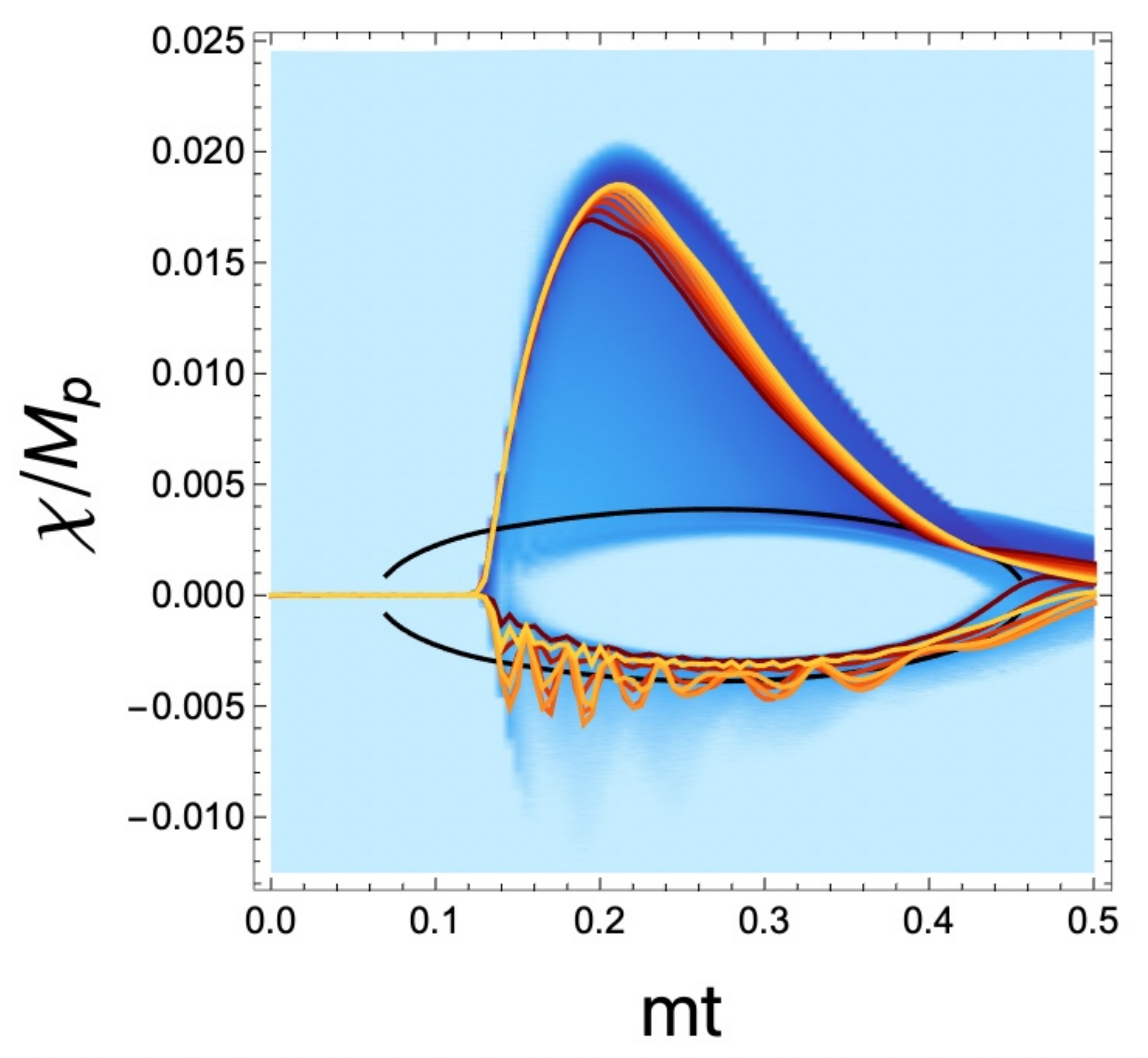}
	\caption{Samples of trajectories following the two different paths. Here the black curve is the local minima of the potential. Note that the lower trajectories with smaller probability of bifurcation in general have more significant oscillations.}
	\label{fig:tracer_traj}
\end{figure}

In Fig.~\ref{f7}, we have plotted the $\chi$ value distributions. In Fig.~\ref{fig:tracer_traj}, we investigate the dynamics of the bifurcation in more details to plot individual trajectories. It is interesting to note that oscillations generically arise in the trajectories with smaller probability, but is absent for the trajectory with larger probability. In position space, this corresponds to oscillations of field values in bubbles, but not outside bubbles.

This effect can be understood as in the more probable trajectory, the kinetic energy drives the trajectory to very large $\chi$ value in a short time, away from the potential minima. This is known as a ``feature scattering'' \cite{Wang:2015kka}. There is thus lack of oscillation since the field is not around the minima. Then, the dynamics is smoother, and the Hubble fraction damps the kinetic energy. 

In the literature, there have been extensive studies on the oscillations and how to extract the information of inflation from these oscillations. For example, these oscillations can behave as primordial standard clocks \cite{Chen:2011zf, Chen:2014cwa} to measure the evolution history of the primordial universe. Our simulation points to an interesting possibility of having primordial standard clock signals in bubbles (such as a void in the large scale structure), but not outside. The oscillations may also bring additional non-Gaussian signatures similar to that of \cite{Sugimura:2012kr}.

This oscillatory behavior of the trajectories also shed light on the dynamics of teh temporary domain walls in multi-stream inflation. Temporary domain walls correspond to the part of position space volume between different preferred trajectories. The dynamics of the domain wall region is interesting. For example, in \cite{Ding:2019mmw}, it is noted that after combination of different inflationary trajectories, the domain wall tension vanishes and the domain wall region starts to expand with the exponential expansion of the universe. This can lead to a broad and smooth intermediate region between the void and outside, which is observationally less constraining from the Sunyaev-Zeldovich effect.

Since the trajectory is oscillating, the temporary domain wall has a time-dependent tension. Since domain walls can leave signatures in density fluctuations \cite{Jazayeri:2014nya}, even if their tension disappears later, it is interesting to explore the observational effects of the oscillatory tension of the domain wall. We hope we can address these issues in a future work.

\section{\label{7}Conclusion}

In this paper we explore some features of multi-stream inflation using lattice simulation. We study a two-field potential with a barrier and find that the trajectories of multi-stream inflation greatly depend on gradient energy. We then analyze in detail how gradient energy play such an important role in the process of bifurcation by comparing the trajectory distribution with or without gradient energy. By comparison we show that gradient energy significantly contributes to the behavior of trajectory distribution. Next, we focus on how the position of this barrier influences bifurcation probability. It turns out that there is a clear error function relation between them. This error function relation cannot be explained simply by initial Gaussian fluctuation, instead it is also related to gradient energy. We also calculate bubble sphericity for different bifurcation probability, confirming that for rare peaks, bubbles tend to be more spherical. Finally, we note that in the less probable trajectories oscillation can take place, corresponding to oscillatory behaviors of observables in bubbles and domain walls, but not outside bubbles. 

With the setup of the lattice code, many other aspects of multi-stream inflation may be studied, such as the details of the curvature power spectrum (which was only estimated by the $\delta N$ formalism), the dynamics of the domain walls and possible impacts with a random potential instead of a simple barrier. We hope to return to some of these topics in future studies.

\begin{acknowledgments}

We are grateful to Jing Liu, Zihan Zhou and Zhuoxin Zhu for very valuable discussion on this article. This work was supported in part by the National Key R\&D Program of China (2021YFC2203100). TC and JJ are supported in part by the NSFC (Nos. 11653002, 11961131007, 11722327, 11421303), by the National Youth Talents Program of China, by the Fundamental Research Funds for Central Universities, by the CSC Innovation Talent Funds, and by the USTC Fellowship for International Cooperation. YW is supported in part by the Hong Kong Research Grants Council fund GRF 16303621 and the NSFC Excellent Young Scientist (EYS) Scheme (Hong Kong and Macau) Grant No. 12022516. All numerical calculations were operated on the computer clusters \textit{LINDA} \& \textit{JUDY} in the particle cosmology group at USTC.

\end{acknowledgments}

\appendix

\section{\label{2}Background Potential\\ without a barrier}

Before simulating multi-stream inflation in a potential with a barrier, we analyze our potential without a barrier, for conventional inflation dynamics without bifurcation. We use a quadratic double field potential to be our background potential
\begin{equation}
	V(\phi,\chi) = \frac{1}{2} m^2 \phi^2 + \frac{1}{2} \alpha m^2 \chi ^2,
\end{equation}
where $m$ is the inflaton mass, $\alpha$ is the $\chi$ term parameter. $\phi_0=16M_p,\chi_0=0$ are the initial values of fields.

For such a simple potential, it is clear that the inflation will proceed very close to single field inflation $V(\phi)=m^2\phi^2/2$. By simulating such a model, we can primarily confirm the validity of our simulation by comparing it with theoretical predictions, what is more important is that it represents the general behavior of our potentials.

\begin{figure}[ht]
    \centering
	\includegraphics[width=0.42\textwidth]{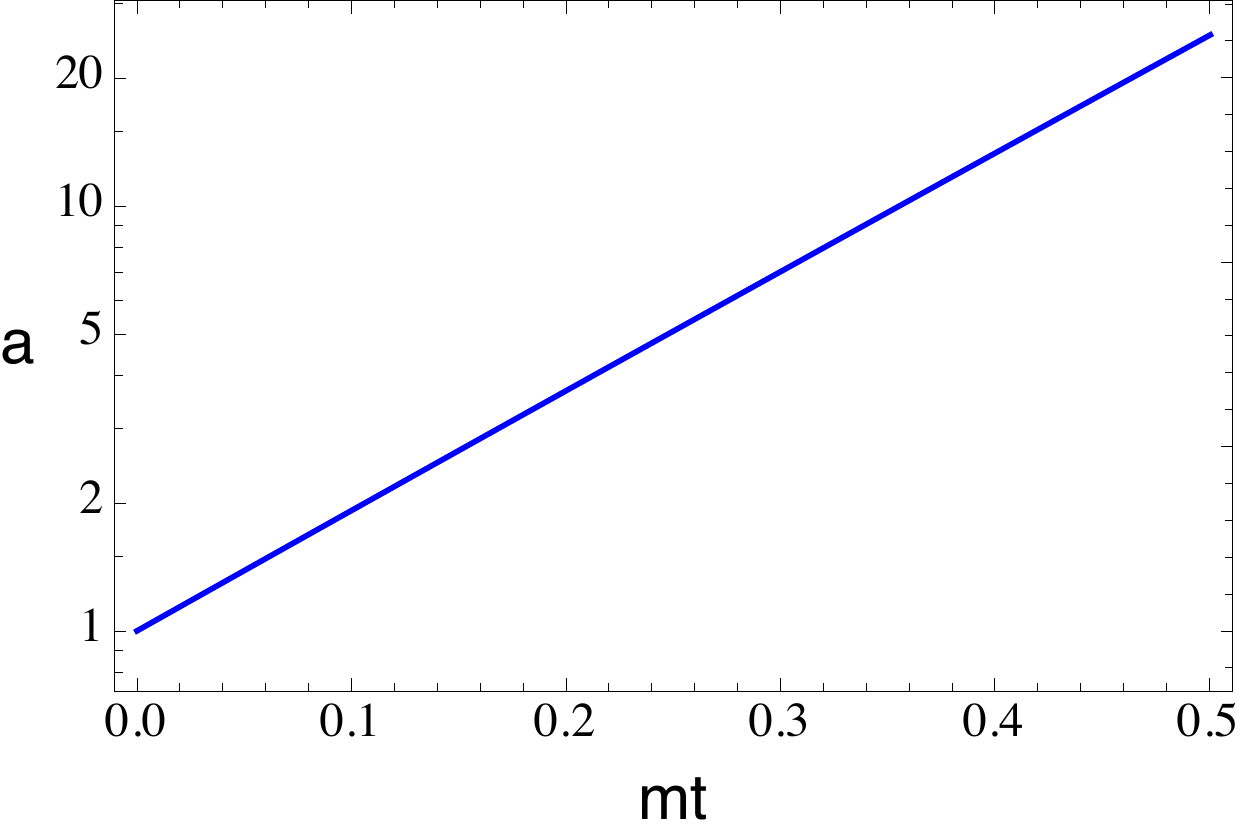}
	\caption{Evolution of scale factor.}
	\label{ap1}
\end{figure}

\begin{figure}[ht]
    \centering
	\includegraphics[width=0.42\textwidth]{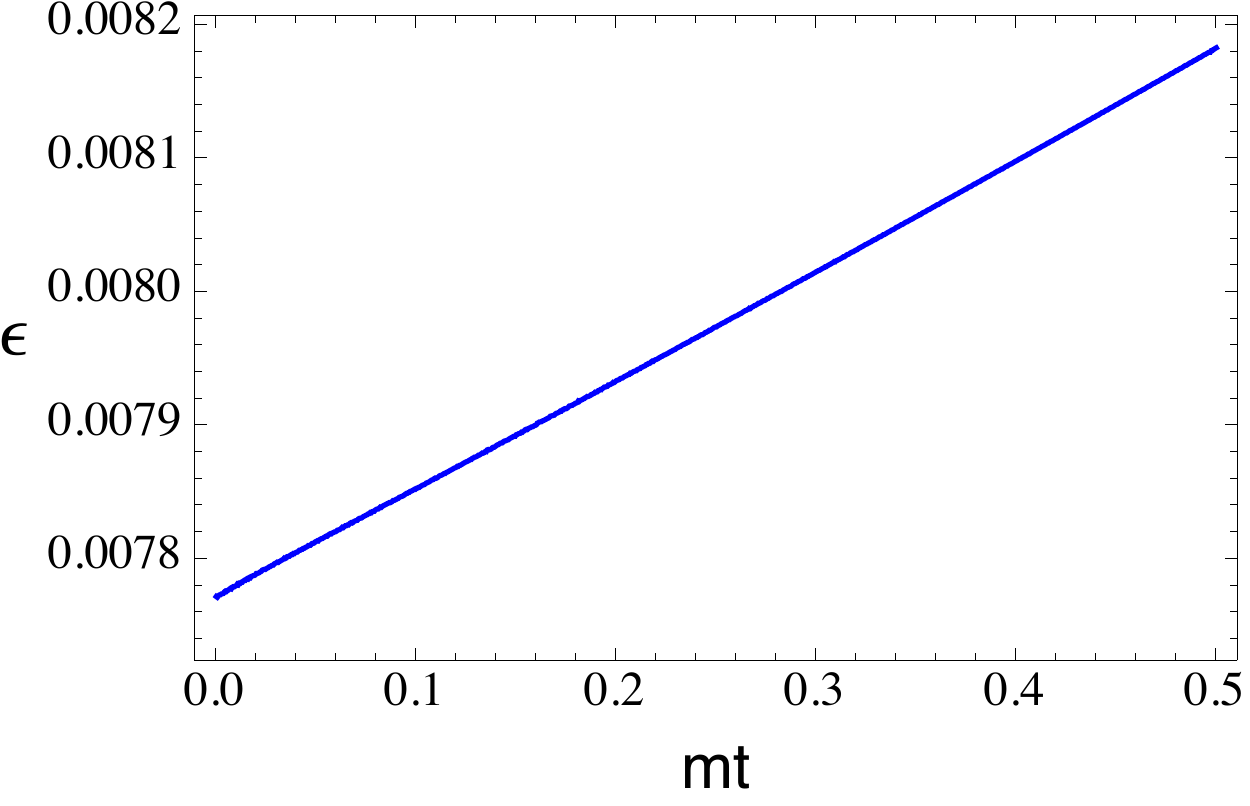}
	\caption{Evolution of slow-roll parameter $\epsilon$.}
	\label{ap2}
\end{figure}

\begin{figure}[ht]
    \centering
	\includegraphics[width=0.42\textwidth]{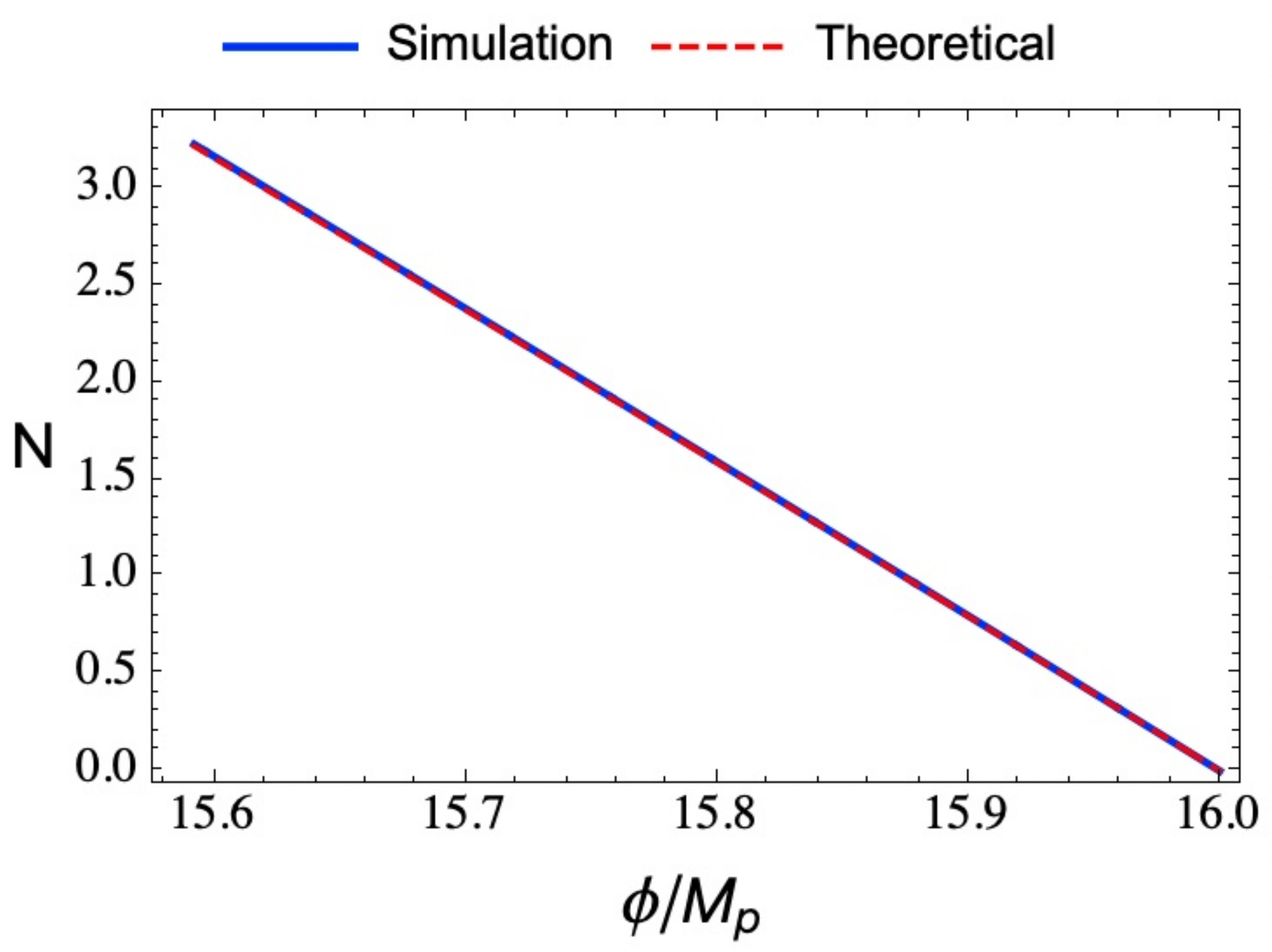}
	\caption{Number of e-folds as a function of $\phi/M_p$, blue curve is simulation result, yellow dashed curve is theoretical result, they match well.}
	\label{ap3}
\end{figure}

\begin{figure}[ht]
    \centering
	\includegraphics[width=0.48\textwidth]{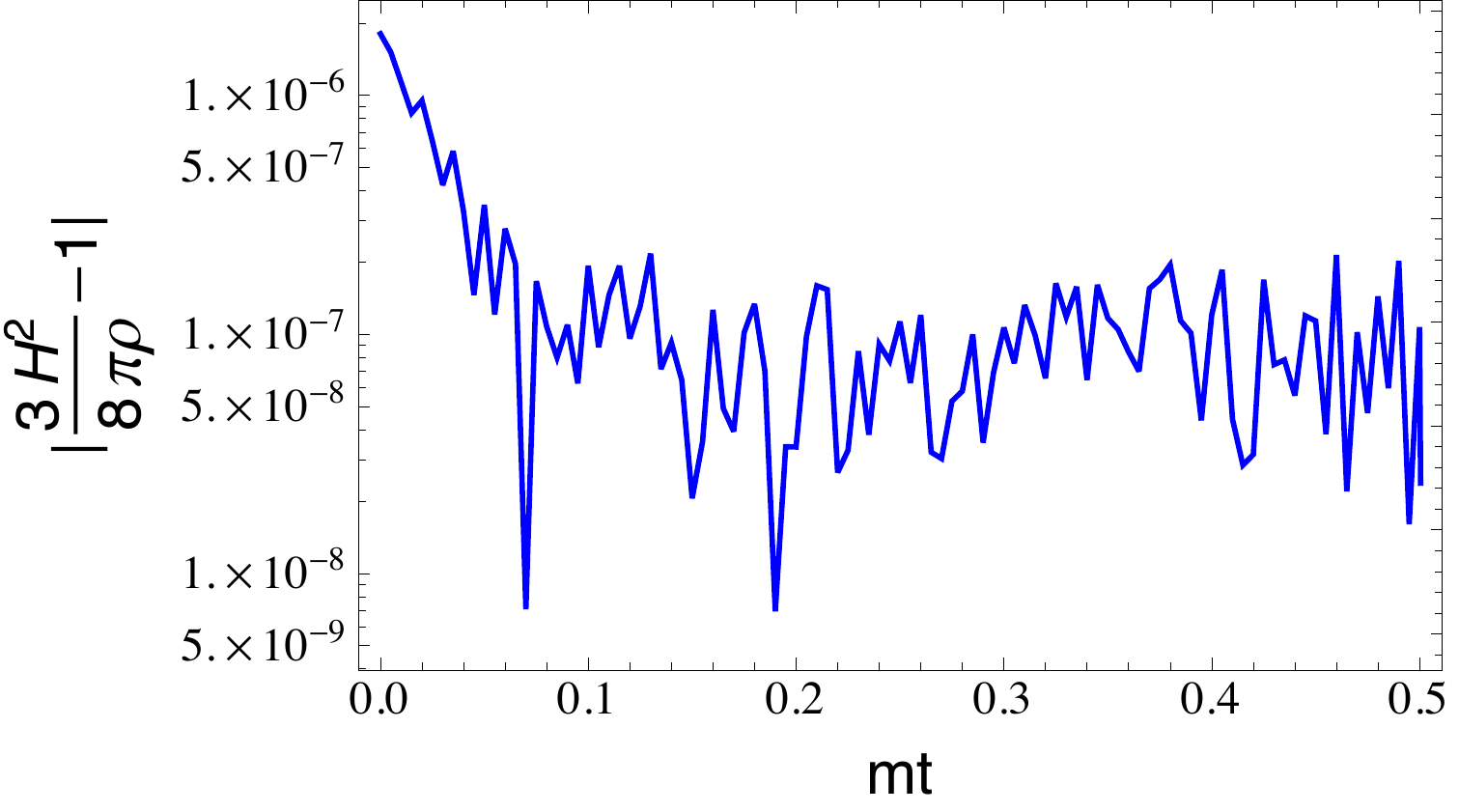}
	\caption{Fractional energy noise as a function of $mt$. It keeps increasing and reach the level of $10^{-7}$.}
	\label{ap4}
\end{figure}

Fig.~\ref{ap1} $\sim$ \ref{ap4} demonstrates some background values during inflation. We can tell that an slow-roll inflation is achieved from the evolution of scale factor and slow-roll parameter. Number of e-folds and value of $\phi$ are also plotted, the blue curve is simulation output while the yellow dashed curve is theoretical prediction, these two curves perfectly match. 

To verify conservation of energy, we pay attention to the fractional energy noise $3H^2/(8\pi\rho) - 1$ during inflation. In our case this quantity is roughly $10 ^ {-7} $ and it tells that the computation is solid. 

It is worth mentioning that when $0<mt<0.1$, the value of $|3H^2/(8\pi\rho)-1|$ is generally decreasing. The reason for it is that gradient energy is not taken into account during initialization, causing such anomaly at the beginning of simulation, yet such level of it is unimportant and barely affects our result.\\

These background values in Fig.~\ref{ap1} $\sim$ \ref{ap4} represent a background inflation, that is to say when we add a barrier to the potential and change related parameters, the curves in are only slightly changed, approximately following the behaviors under a standard inflation. This is confirmed by actual simulation.

\section{\label{a1}Sphericity of Lattice Bubble}

In Sec.~\ref{5} we extract bubbles from the lattice and calculate sphericity. Here the challenge is to define sphericity on a lattice simulation. Since we have limited lattice resolution, we need the definition to be robust with relatively low lattice resolution. For this purpose, we define the sphericity as follows.

\begin{figure}[ht]
    \centering
	\includegraphics[width=0.42\textwidth]{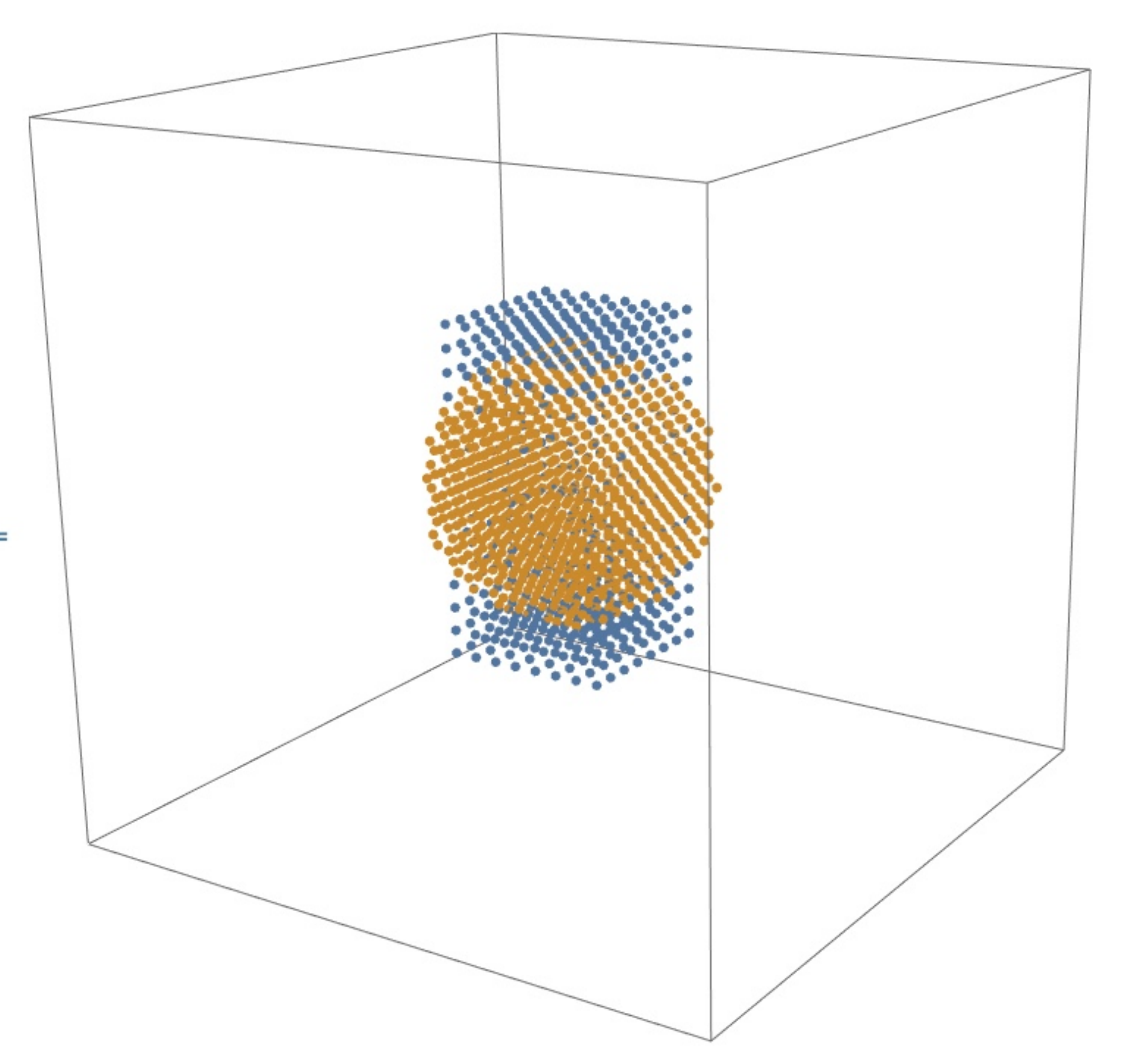}
	\caption{Calculation of sphericity. We want to calculate the sphericity of the object with blue points. Sphere with yellow points is the reference sphere that we take. Sphericity is then the ratio of number of mutual points that object and reference sphere share to number of points of object.}
	\label{fsp1}
\end{figure}

For an object consisting of $N_t$ lattice points, we average all the points separately for three coordinate directions to obtain the center point of this object, then we search the lattice points and select $N_t$ points closest to the center point to form a reference sphere. We define $N_m$ as the number of mutual points that object and reference sphere share, then
\begin{equation}
	\text{Sphericity}=\frac{N_m}{N_t}.
\end{equation}
Fig.~\ref{fsp1} is an example of such sphericity calculation.\\

\begin{figure}[ht]
    \centering
	\includegraphics[width=0.47\textwidth]{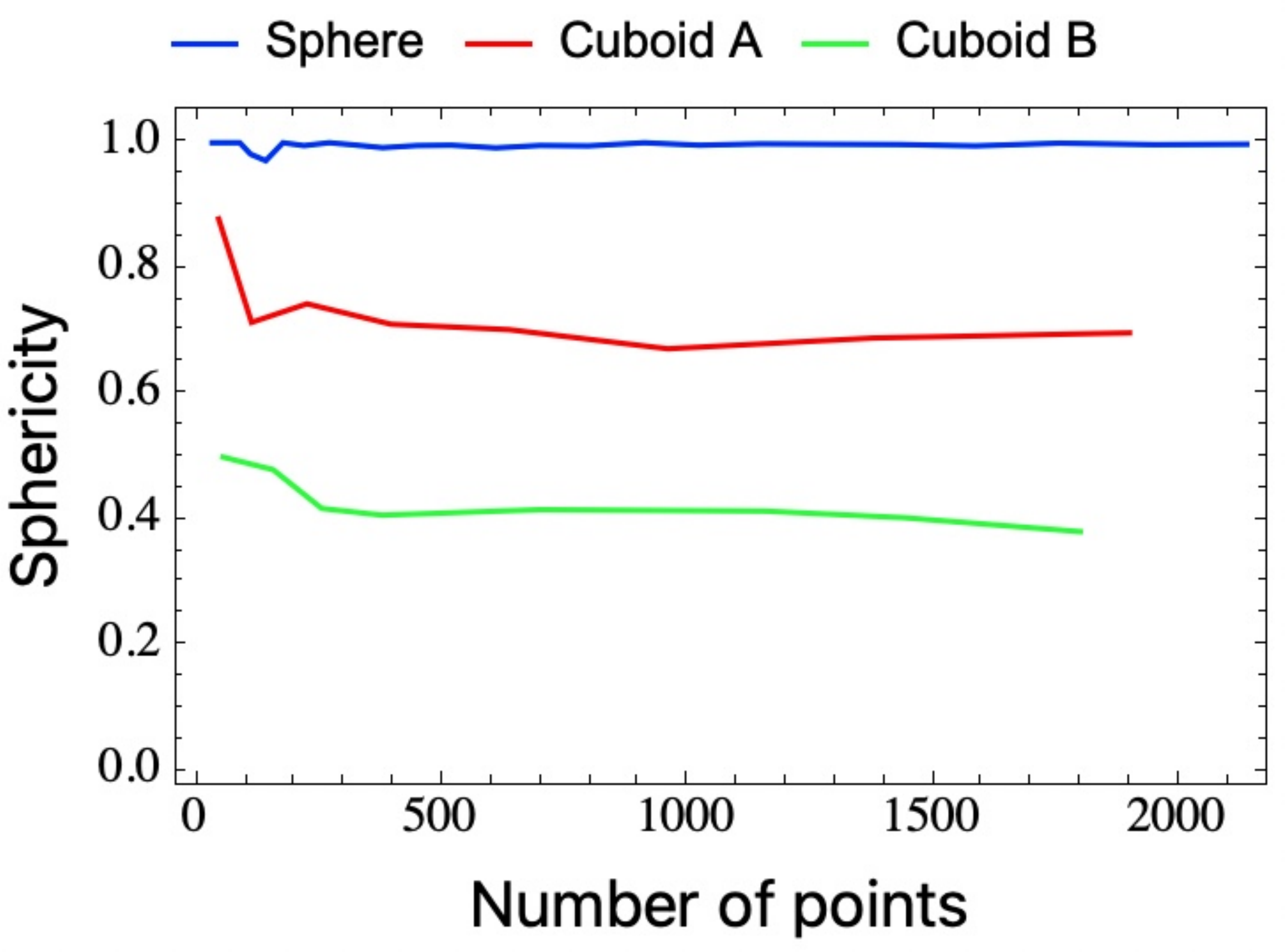}
	\caption{Sphericity as a function of number of points of the calculated object. Different colors represent different shapes of objects.}
	\label{fsp2}
\end{figure}
 
In Fig.~\ref{fsp2} we calculate the sphericity of objects consisting of different number of lattice points. It turns out that for different objects the sphericity is stable. We set 300 as an threshold, that is to say we only take into account the sphericity of objects consisting of more than 300 lattice points. Our results in \ref{5} shows obvious tendency of sphericity, that is another confirmation of validity of this definition of sphericity.

\bibliography{multi_stream_bubble}

\end{document}